From behrndt@physik.hu-berlin.de Wed Sep  2 15:05:17 1998
Date: Wed, 2 Sep 1998 13:43:15 +0200 (MET DST)
From: Klaus Behrndt <behrndt@physik.hu-berlin.de>
To: Klaus Behrndt <behrndt@physik.hu-berlin.de>
Subject: file

\documentclass[12pt]{article}
\providecommand{\href}[2]{#2}   
\usepackage{graphicx}
\usepackage[normal,small,bf]{caption}
  \newcommand{\be}[3]{\begin{equation}  \label{#1#2#3}}
  
\newcommand{\ee}{ \end{equation}}
\newcommand{\ba}{\begin{array}}
\newcommand{\ea}{\end{array}}

\renewcommand{\arraystretch}{1.7}
\let\LARGE=\Large
\let\Large=\large

 \def\unit{\hbox to 3.3pt{\hskip1.3pt \vrule height 7pt width .4pt \hskip.7pt
\vrule height 7.85pt width .4pt \kern-2.4pt
\hrulefill \kern-3pt
\raise 4pt\hbox{\char'40}}}

\begin{document}


\thispagestyle{empty}
\rightline{HUB-EP-98/53}
\rightline{hep-th/9809015}

\vspace{1.5truecm}

\centerline{\bf \LARGE AdS gravity and field theories at fixpoints}
\vspace{2truecm}

\centerline{ {\bf Klaus Behrndt} \footnote{e-mail:
  behrndt@physik.hu-berlin.de\\ 
  Based on a talk/poster presented at
  the ``Superfivebranes and Physics in 5+1 Dimensions'' workshop,
  Trieste, Italy, April 1998 and at ``Strings98'' conference, ITP,
  Santa Barbara, USA, June 1998. 
   } }
\vspace{.5truecm}

\centerline{\em Humboldt University Berlin}
\centerline{\em Invalidenstrasse 110, 10115 Berlin, Germany}

\vspace{1.2truecm}


\vspace{.5truecm}

\begin{abstract}
\noindent
The renormalization group flow of the worldvolume theory depends very
much from the number of unbroken supersymmetries. In the dual $AdS$
picture we break supersymmetry by adding different types of BPS black
holes.  We argue, that this BPS black hole causes a non-trivial
renormalization group flow in the worldvolume field theory and
especially a regular horizon translates into a non-trivial IR
fixpoint. For this interpretation we have to rewrite the $AdS$ models
into a flat space description with a linear dilaton vacuum. The dual
models (linear dilaton and the $AdS$ vacuum) can be seen as the
different sides of a domain wall. We discuss the cases of $AdS_3$ and
$AdS_5$.

\vspace{1cm}

\end{abstract}


\newpage


\section{Introduction}

The AdS/CFT correspondence provides an interesting framework to relate
super Yang-Mills theories to supergravity theories.  However, in a
low-energy approximation both theories exhibits singularities, for
a bulk observer the supergravity solutions are typically singular and
a worldvolume observer has to deal with IR or UV singularities which
are often non-renormalizable. These singularities are an indication,
that with a certain energy scale a strong interaction between the
worldvolume and bulk degrees appears, which cannot be neglected
anymore.

{From} the supergravity point of view these singularities could just
mean that a hidden space opens up.  An interesting observation is,
e.g., that all singular branes become regular in a dual, conformally
rescalled, frame \cite{080}.  And reducing the spherical part of this
dual frame one reaches the $AdS$ vacuum of a domain wall solution,
where the singularity of the original brane indicates the appearance
of the second (asymptotically flat) vacuum.  Below we will find an
analogous picture for 3-d black holes.

However, a subclass of branes are regular and for these cases we may
expect that the worldvolume and bulk theories decouple
consistently. Since there are no singularities we can choose the
parameters in a way that the perturbation theory is under control and
we can work in a semiclassical approximation, i.e.\ $N \rightarrow
\infty$ (large number of branes) and $\alpha' \rightarrow 0$.  These
non-singular branes appear in various dimensions and have been
classified in \cite{120}. They have in common that: (i) near the
horizon the space time factorizes into $AdS_p \times S_q$ and (ii)
they are scalar free. As a side remark, in this set of solutions there
is only one brane, that does not appear in standard compactification
of string or $M$-theory - this is the self-dual membrane in 8
dimensions\footnote{ The 8-d metric is $ds^2 = H^{-2/3} dx_{\|} +
H^{2/3} dx_{\perp}$.} (it is also not a BPS configuration).

The anti deSitter space is asymptotically not flat (non-trivial
boundary) and it is known for long time that the $AdS$ isometry group
is realized as conformal group on this boundary. This led to the
conjecture \cite{050}, \cite{260} that the boundary CFT is dual to the
superconformal field theory on the worldvolume.  A great deal of
attention received the odd cases.  E.g.\ the boundary theory of
$AdS_7$ is expected to be dual to (non-critical) string theory
describing the worldvolume of the $M$5-brane; or of $AdS_5$ should be
dual a 4-d super Yang-Mills and the boundary theory of $AdS_3$ is a
2-d $\sigma$-model.

But does the brane really reside at the infinite boundary of the anti
deSitter space? As recently stressed in an FAQ by Maldacena \cite{053}
this point of view has to be taken with care.  
The better way of thinking is that the brane is everywhere, i.e.\ the
projection on any radial hypersurface is dual to the worldvolume field
theory. In this interpretation the radial coordinate in supergravity
sets the energy scale in the worldvolume description, e.g.\ the
infinite boundary describes the UV region.  Since at this boundary the
$AdS$ group acts as (super) conformal group we have to expect that the
world volume theory has an UV-fixpoint of renormalization group (RG)
equation.

\begin{figure}
\begin{center}
\includegraphics[angle=0, width=130mm]{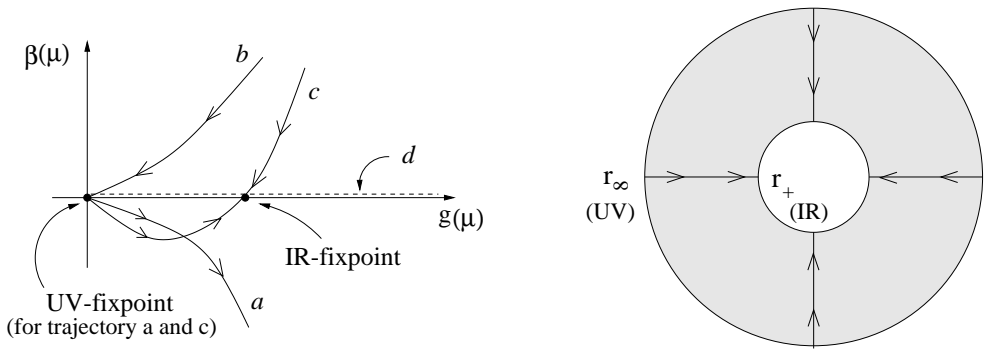} 
\end{center}
\caption{In the left figure, the arrows indicate the IR scaling
behaviour of the RG $\beta$-function.  Case $a$: There is no
IR-fixpoint. At large distances the coupling constant $g$ diverges,
but it is UV-free.  Case $b$: Is the opposite, in the UV limit
the coupling diverges whereas in the infrared the model is free. Case
$c$: This is a model with two fixpoints and finally case $d$ shows a
model where the $\beta$ function vanishes identical.  The right figure
shows the corresponding $AdS$ picture, where the different cases are
related to singular or regular behaviour near the inner boundary
(horizon). }
\label{beta} 
\end{figure}

But, what happens if we enter the $AdS$ bulk? Since the radial
distance sets the energy scale, this translates into a RG flow
departing from the UV fixpoint. The subsequence behaviour depends very
much on the amount of unbroken supersymmetries.  If we have a
sufficient amount of supersymmetry we may expect that the conformal
symmetry will survive also at lower energies.  E.g.\ $N$=4 super Yang
Mills is conformal for all energies, the RG $\beta$-function vanishes
identical.  But already if one has ``only'' $N$=2 supersymmetry the
scaling symmetry is generically broken.  As shown in figure \ref{beta}
different scenarios are possible. First, nothing happens like for
$N$=4 super Yang Mills, i.e.\ the $\beta$ function vanishes identical
and the model is scale invariant everywhere (case $d$).  A second
possibility is that we move towards a strongly coupled region with no
IR fixpoints as in case $a$. The case $b$ shows exactly the opposite,
where we have no UV but an IR fixpoint. Both cases appear naturally in
$N$=2 super Yang Mills. Finally for $N$=1 super Yang Mills we can
expect the behaviour shown in case $c$, i.e.\ starting from a UV
fixpoint the RG flow goes towards an IR fixpoint, where the scaling
symmetry is restored.

What does this fixpoint behaviour mean for the $AdS$ picture?  An
obvious possibility to implement this behaviour is to consider a
non-trivial background, which ``creates'' a second boundary of the
$AdS$ space and allows unbroken supersymmetries. Again, on this second
boundary we may expect that the $AdS$ group is realized as conformal
group and therefore, the worldvolume field theory approaches a
non-trivial fixpoint. As candidates for this scenario we may take BPS
black holes. The asymptotic vacuum translates into the UV fixpoint and
moving towards the horizon we approach the IR region of the
worldvolume field theory.  In the case that the horizon is regular,
the moduli will move towards a finite value causing an interacting
worldvolume theory at a non-trivial IR fixpoint. On the other hand if
the horizon is singular the coupling constant will diverge. This does
not necessarily mean that the model does not make sense - it could
simply indicate that we have to change the variables.  E.g.\ for the
$AdS_5$ case: BPS black holes of $N$=4 gauged supergravity should
describe RG flow of $N$=2 super Yang-Mills whereas BPS black holes of
$N$=2 gauged supergravity correspond to $N$=1 super Yang-Mills.
Notice, it is important to consider supersymmetric black holes, the
Schwarzschild deSitter black holes will break all supersymmetries and
are usually interpreted as adding temperature to a given super Yang
Mills theory (with a compact Euclidean time).

One may object that the horizon is not really a boundary in the
supergravity sense, by a change of coordinates we can extend the
spacetime beyond the horizon. But the same is also true for the field
theory, which has also a realization beyond the (non-trivial)
fixpoint. However, the system is trapped between two fixpoints under
the RG flow. In order to cross the horizon one has to change the
coordinate system (e.g.\ Kruskal coordinates), which translates into
field theory in a change of operators, especially the Hamiltonian
describing the evolution in a given time has to be changed.  But when
formulated in the asymptotic flat coordinate system it is the maximal
extendable region, see also \cite{270}.

It is the aim of this letter to discuss this scenario in more detail.
We will start with the $AdS_3$ case and investigate the outer region
of BTZ black hole \cite{160}. In order to extract the different CFTs
it is natural to employ the Chern-Simons formulation of 3-d gravity
\cite{170}, \cite{180}, which makes the holographic nature manifest.
It is also worth to mention that in this setup it is important that
the BTZ black hole represents a discrete identification along one of
the $AdS$ Killing vectors. This construction devides the
$AdS$ space in different domains \cite{190} and our annulus is one of
them. Furthermore, we will show in section 2.1 that the BTZ black hole
can be recast into a domain wall interpolating between an asymptotic
$AdS$ vacuum and a flat space with a linear dilaton. 
Using the Chern-Simons setup for the $AdS$ vacuum we will give a
derivation of the CFTs in section 2.2.  Finally, in section 3 we turn
to the $AdS_5$ case, where the situation is much more involved. We
will focus on the question how we can break supersymmetry by adding
various types of BPS black holes, corresponding to $N$=2 or $N$=1
super Yang Mills as worldvolume theory.


\section{The joy at $AdS_3$}

A good starting point is to describe the situation for $AdS_3$ gravity
and we are interested in a background with the 2-d spatial geometry given
by an annulus. A natural candidate for this is the outer region of the
BTZ black hole \cite{160}. Before we discuss the CFTs near the
boundaries it is usefull to discuss the black hole and their
dual string (=domain walls) as solutions of 3-d gravity.

\subsection{3-d gravity: black holes and domain walls}

The BTZ black hole is given by
\be100
ds^2 = - e^{-2V(r)} \ dt^2 + e^{2V(r)} \ dr^2 + \Big({r\over l}\Big)^2 \
        \Big( dy -\frac{r_- r_+}{ r^2} \, dt  \Big)^2
\ee
with
\be110
e^{-2 V(r)} = {(r^2 - r_-^2)(r^2 - r_+^2) \over r^2 l^2}
\ee
The horizons of the BTZ black hole are located at $r=r_{\pm}$, the
mass and angular momentum are given by $M = {r_+^2 + r_-^2 \over
l^2}$ and  $J= {2 r_+ r_- \over l^2}$.  It solves the
equations of motion coming from the three-dimensional action
\be120
S = \frac{1}{2 \kappa_3^2} \int\ d^3 x \ \sqrt{-g} \
             (R \ + \ {2 \over l^2} )
\ee
where the cosmological constant $l$ is given by the $AdS$ radius.
Since we are in 3 dimensions we can dualize the cosmological constant
to an antisymmetric tensor
\be130
S = \frac{1}{2 \kappa_3^2} \int\ d^3 x \ \sqrt{-g} \
             (R \ - \ {1 \over 12} H^2 )
\ee
with
\be140
H^{\mu\nu\rho} = {2 \over l}\, {1 \over  \sqrt{g}} \epsilon^{\mu\nu\rho}
 \qquad \qquad 
(B_{0y} = {r^2 \over l^2}) \ .
\ee
This indicates that there is a dual domain wall solution,
which is a string in 3 dimensions. In fact $T$-dualizing the
$y$ direction one obtains
\be150
\ba{l}
ds^2 = \left( {r_-^2 + r_+^2 \over l^2} - {r_-^2 r_+^2 \over l^2 r^2}
\right) \, dt^2 - 2 \, dt dy + {l^2 \over r^2} dy^2 + e^{2V} dr^2
\ , \\ 
e^{-2\phi} = {r \over l} \qquad , \qquad B = {r_- r_+ \over r^2}
dt \wedge dy \ .
\ea
\ee
Depending on the extreme or non-extreme case one can simplify this
solution further; see \cite{200}, \cite{220}.  For the non-extreme case,
we define new coordinates by
\be160
t \rightarrow { l \over \sqrt{r_+^2 - r_-^2}} ( y - t)
\quad , \quad y \rightarrow {1 \over l \sqrt{ r_+^2 - r_-^2}} 
( r_-^2 y - r_+^2 t ) \quad , \quad
r^2 \rightarrow r^2 + r_-^2
\ee
and find (after a gauge transformation in $B\rightarrow 
B - {r_- \over r_+}$)
\be170
ds^2 = {1 \over H}\left[- (1 - {\mu \over r^2}) dt^2 + dy^2 \right]
+ {l^2 \over r^2}  {dr^2 \over 1 - {\mu \over r^2}} \quad , \quad
e^{-2\phi} = {r^2 \over l^2} \, H \quad , \quad
B = {r_- \over r_+ H} dy \wedge dt
\ee
with $H = 1 + {r_-^2 \over r^2}$. It is interesting to note, that
by a further redefinition of the radial coordinate $r^2 = \mu \cosh^2 
\lambda/l$ we get the metric
\be172
ds^2 = - {r_+^2 - r_-^2 \over r_+^2 \coth^2 \lambda/l - r_-^2} dt^2 \ +\  
d \lambda^2 \ +\  {(r_+^2 - r_-^2) \coth^2 \lambda/l 
\over r_+^2 \coth^2 \lambda/l - r_-^2} d y^2
\ee 
which after compactification over $y$ is exactly the 2-d black hole
discussed as in \cite{210}. There is however an important difference
to this exact 2-d string background, namely the non-trivial
antisymmetric tensor, which becomes a gauge field upon dimensional
reduction.  Near the boundary at $\lambda = 0$ however the $B$ field
drops out and the solution coincides with the known exact string
background. 

For the extreme case ($r_-^2 = r_+^2 = r_0^2$) we choose
the coordinates
\be180
t \rightarrow {l^2 \over 2 r_0^2} v \qquad , \qquad 
y \rightarrow {1 \over 2} \left(v - {2 r_0^2 \over l^2} u \right) 
\qquad , \qquad
r^2 \rightarrow r^2 + r_0^2
\ee
and obtain
\be190
ds^2 = { 1 \over H} \left[ du dv + {r_0^4/l^2 \over r^2}
du^2 \right] + {l^2 \over r^2} \, dr^2 \quad , \quad 
e^{-2 \phi} = {r^2 \over l^2} H \quad , \quad
B= {1 \over 2 H } du \wedge dv
\ee
with $H = 1 + {r_0^2 \over r^2}$.   Note, in order to
go from the non-extreme to the extreme string solution one has first
to perform the Lorentz rotation that is included in the transformation
(\ref{160}) and afterwords make the standard extreme limit.

In addition, the BTZ black hole itself can be written as a
string solution.  Defining a new radius in the extreme case ($r_+ =
r_- = r_0$) by\footnote{The corresponding transformation for the
non-extreme case can be found \cite{020}.}
\be200
r^2 \rightarrow r^2 + r_0^2
\ee
the BTZ black hole (\ref{100}) becomes
\be210
ds^2 = {r^2 \over l^2} \left[du dv + {r_0^2 \over r^2} du^2 \right]
 + {l^2 \over r^2} dr^2
\ee
where $v/u = y \pm t$. Both (extreme) string solutions have an
important difference: they corresponds to different asymptotic
vacua. The solution given in (\ref{190}) is asymptotically flat
with a linear dilaton background
\be220
ds^2 = du dv + {l^2 \over r^2} dr^2 = du dv + d \lambda^2
\quad , \quad \phi = - {1 \over l} \, \lambda
\ee
where the background charge is given by the $AdS$ radius. 
Hence, in the asymptotic vacuum ($ \lambda \rightarrow \infty$)
we are in a weakly coupled region. On the other hand the
dual string in (\ref{210}) is asymptotically anti-de Sitter
\be230
ds^2 = {r^2 \over l^2} dudv + {l^2 \over r^2} dr^2 \ .
\ee
with vanishing (or fixed) dilaton. Anti-deSitter gravity in 3
dimensions exhibits holography, which becomes manifest if one
formulate it as a topological Chern-Simons model.  As consequence,
also the dual model with a linear dilaton background has to show
holography, see also \cite{230}.

\begin{figure}
\begin{center}
\includegraphics[angle=0, width=120mm]{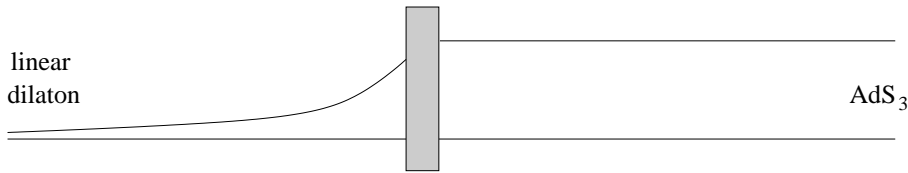} 
\end{center}
\caption{It shows the domain wall solution. On one side
we have the $AdS_3$ vacuum and the solution is given by
(\ref{210}). On the other side we have a linear
dilaton vacuum and the solution corresponds to (\ref{190}).
The coupling constant is fixed on the $AdS$ side whereas it
vanish in the linear dilaton vacuum. Both sides are
dual to each other.
}
\label{domainWall} 
\end{figure}

Notice, near the core ($r \simeq 0$) both solutions (\ref{190}) and
(\ref{210}) are equivalent and we may identify and see them as
as two sides of a domain wall, see figure \ref{domainWall}.

\subsection{Conformal field theories of $AdS_3$ gravity}


Let us come back to the $AdS$ side in figure 2 and let us determine
the CFTs.  It is known that Einstein-anti-de Sitter gravity in $2+1$
dimensions as given by the action (\ref{120}) is equivalent to a
Chern-Simons theory \cite{170}, \cite{180}. Choosing conventions where
the three-dimensional gravitational coupling is related to the level
$k$ by
\be600 
k =\frac{2 \pi l }{\kappa^2_3} 
\ee 
and decomposing the diffeomorphism group $SO(2,2) \simeq SL(2,{\bf
R})_L \times SL(2, {\bf R})_R$ the 3-dimensional action can be written
as
\be610
S = S_{CS} [A] \ - \  S_{CS} [\bar A]
\ee
with
\be620
S_{CS} [A] = \frac{k}{4 \pi} \int_{M_3} d^3 x
        {\rm Tr} \ (AdA + \frac{2}{3} A^3 ) \ .
\ee
The gauge field one-forms are
\be630
A = (\omega^a + \frac{1}{l} e^a) \ T_a \ \in SL(2,{\bf R})_R \ , \hspace{1cm}
\bar A = ( \omega^a - \frac{1}{l} e^a) \ \bar T_a \ \in SL(2,{\bf R})_L.
\ee
where $\omega^a \equiv \frac{1}{2} \epsilon^{abc} \omega_{bc}$ are
given by the spin-connections $\omega_{bc}$ and $e^a$ are the
dreibeine.  Considering non-trivial boundaries the Chern-Simons theory
is not invariant under gauge transformations and as a consequence
gauge degrees of freedom do not decouple and become dynamical on the
boundaries. These are the degrees of freedom of the conformal field
theories living at the boundaries.

In the following we will discuss this procedure for the BTZ black hole.
The geometry of the manifold is $M_3 = {\bf R} \times \Sigma$, where
${\bf R}$ corresponds to the time of the covering space of $AdS_3$ and
$\Sigma$ represents an ``annulus'' $r_+ \leq r < \infty$.

Calculating the gauge connections $A= A^a T_a$ and $\bar A =
\bar A^a \bar T_a$  for the BTZ solution (\ref{100})
one finds (for details see \cite{020}, \cite{240})
\be640
\begin{array}{l}
A = \Big( e^{-V} \, T_0  + {r \over l} (1 - {r_- r_+ \over r^2}) \, T_1 \Big)
        {dv \over l} + e^{V} (1 + {r_- r_+ \over r^2})\, T_2  \
        \frac{dr}{l} \ , \\
\bar A = \Big( e^{-V} \, T_0  - {r \over l} (1 + {r_- r_+ \over r^2}) \,
        T_1 \Big) {du \over l}  - e^{V} (1 - {r_- r_+ \over r^2}) \,
        T_2 {dr \over l} \ .
\end{array}
\ee
The corresponding gauge fields strengths for these gauge potentials
vanish, i.e.\ they represent only gauge degrees of freedom\footnote{For
the $SL(2, {\bf R})$ generators we choose the represention
\renewcommand{\arraystretch}{1.0}
\[
T_0 = \frac{1}{2} \ 
\left (
   \begin{array}{cc}
   0     & -1   \\
   1     & 0  \\
\end{array}
\right ), \ \ \ \
T_1 = \frac{1}{2} \ 
\left (
   \begin{array}{cc}
   0     & 1   \\
   1     & 0  \\
\end{array}
\right ), \ \ \ \
T_2 = \frac{1}{2} \ 
\left (
   \begin{array}{cc}
   1    & 0   \\
   0    & -1  \\
\end{array}
\right ).
\]
}
\be650
\ba{l}
A = g^{-1} dg \qquad {\rm with:} \quad g = e^{\lambda T_2} e^{v T_1}\ ,
\qquad \sinh \lambda = \, {l \over r_+ - r_-} \, e^{-V} \\
\bar A = \bar g^{-1} d\bar g \qquad {\rm with:} \quad 
\bar g = e^{\lambda T_2} e^{u T_1} \ ,
\qquad \sinh \lambda = \, - {l \over r_+ + r_-} \, e^{-V} \ .
\ea
\ee
In order to extract the CFTs at the boundaries we have to perform two
steps: (i) we have to add boundary terms that impose the correct
boundary conditions and (ii) we have to mod out the isometry
group of the BTZ solution (corresponds to  the Killing
direction that has been periodically identified in constructing
the BTZ black hole).

(i) As dictated by the Chern-Simons solution (\ref{640}) we will impose
as boundary conditions
\be660
A_u = \bar A_v =0
\ee
and therefore we add as boundary term to the action
\be670
\delta S = {k \over 8\pi} \int_{\partial M} Tr
(A_v A_u + \bar A_v \bar A_u) \ .
\ee
Since we have flat gauge connection we insert $A = g^{-1} dg$ and
$\bar A = \bar g^{-1} d \bar g$ into the action and one obtains as
result an $SL(2, {\bf R})$ WZW model \cite{150}. 

(ii) The subgroup that has to be modded out can be determined from the
Chern-Simons fields (\ref{640}): for $r \rightarrow \infty$ one finds deformation
along $T_{\pm}$ group direction and near the horizon deformations along
the spatial $T_1$ direction.  Notice, that the degrees of freedom of the
boundary CFT are the ``broken gauge degrees of freedom'' (that become
dynamical on the boundary) and the isometries represent residual
symmetries.  To get the right CFT we have to mod out the residual
symmetries and hence obtain gauged WZW models.  It has been discussed
some time ago, that by gauging a lightcone group direction one truncates
the $SL(2, {\bf R})$ to a Liouville model and by gauging the spatial
direction one obtains a 2-d black hole solution (see \cite{140},
\cite{210}), which becomes a string background in the extreme limit.



These CFTs coincides with our expectation from the domain wall discussion,
i.e.\ the Liouville model corresponds to the linear dilaton vacuum,
whereas the domain wall itself is described by a 2-d black hole
(\ref{172}) in the non-extreme case or the string solution (\ref{190})
in the extreme case.  Both solution are known to be exact CFTs.
In this approach we do not consider the BTZ black hole as small
perturbation around the asymptotic vacuum, but as interpolating
solution between two CFTs. Due to the holographic nature, the complete
bulk physics will be fixed by these boundary CFTs.  

Therefore, the outer-CFT is
an ${SL(2, {\bf R}) \over SO(1,1)}$-WZW model (Liouville model)
defined by the 2-d action \cite{210}, \cite{020}
\be240
S = {k - 2 \over 4 \pi l^2 } \int \left[ (\partial \lambda)^2 - 
Q l \lambda R^{(2)}
+ \mu e^{-2 \lambda/l } \right]
\ee
with the level of the WZW model $k = {l \over 4 \alpha'}$ and the
background charge $Q ={k -1 \over k - 2}$.  The Liouville field
$\lambda$ describes radial fluctuations: $e^{-2 \lambda/l} = \Big({l
\over r}\Big)^2$ and the central charge of this model is
\be250
c_{outer} = 1 + 6(k-2) Q^2 = {3k \over k-2} - 2 + 6k \ .
\ee
In the classical limit of large $AdS$ radius ($k \rightarrow \infty$)
only the last term contributes. 

The inner-CFT is the standard ${SL(2, {\bf R})
\over U(1)}$ gauged WZW model \cite{140}, \cite{210}
with the central charge
\be260
c_{inner} = {3k \over k-2} - 1 \ .
\ee
The Lagrangian of this CFT is given by a 2-d $\sigma$-model with
background fields given in (\ref{170}) for the non-extreme case or in
(\ref{190}) for the extreme case.  The difference in the central
charges indicates that there is no exact marginal deformation
connecting the outer and inner CFT. However, let us stress that at any
finite point in space time one can promote the background to an exact
CFT. On one hand the gauged WZW model can be made exact by changing
the renormalization group scheme (field redefinitions) \cite{320} and
on the other hand the BTZ black hole is locally at any point
$AdS_3$. The different central charges indicate the non-trivial
global structure of the model and it is better thought of as an
interpolating solution between two (different) CFTs.












\section{The worry with $AdS_5$}

The $AdS_3$ case was especially simple, mainly because we were dealing
with 2-d CFTs, but also, because we did not consider any matter.  One
may motivate this, because in 3 dimensions gauge fields are dual to
scalar fields and since $AdS_3$ is typically discussed as the near
horizon region of strings all scalars are fixed (they become constant
near regular horizon).  

For $AdS_5$ the situation is much more complicated, not only that we
are dealing with 4-dimensional CFTs, but also because we have no
reason to ignore matter. Instead we want to use non-trivial gauge
fields to break successive supersymmetry.  Following the picture
described in the introduction we will consider charged BPS black holes
that breaks supersymmetry and tries to verify the different cases as
shown in figure \ref{beta}.  In this setup the number of unbroken
supersymmetries is related to the number of independent charges.

When viewed from a 10-dimensional perspective every charge corresponds
to one brane and at least some of them have to intersect in order to
break supersymmetry.  The single 3-brane, e.g., gives the $AdS_5$
vacuum. It breaks half of the 10-d supersymmetry and corresponds thus
to $N$=4 super Yang Mills (case $d$ in figure \ref{beta}). Upon
compactification over the spherical part, the 3-brane charge enters
the cosmological constant of the $AdS$ space.  By adding a further
brane we break again one half of supersymmetry, i.e.\ for the
worldvolume theory we expect to get $N$=2 super Yang Mills. In order
to reach $N$=1 super Yang Mills we have to consider 3 and 4 charge
configurations (typically both cases have the same amount of unbroken
supersymmetries).

So, from the 5-dimensional point of view we have to discuss black holes
with 1, 2 or 3 independent charges (one charge is absorbed by the
cosmological constant). Thus, we consider the $STU$-model 
for gauged supergravity given by the action
\be400
S \sim \int \left[ {R \over 2} + g^2 V(X) - {1 \over 4}
G_{IJ} F^{I}_{\mu\nu} F^{J\, \mu\nu} - {1 \over 2} G_{IJ} \partial_{\mu}
X^I \partial^{\mu} X^J \right] + {1 \over 48}
\int F^1 \wedge F^2 \wedge A^3
\ee
with $(S,T,U) = X^I= (X^1, X^2, X^3)$ which have to fulfil the
the constraint $STU =1$. The potential and the scalar metric is
\renewcommand{\arraystretch}{1}
\be410
V(X) = 9 \; V_I V_J ( X^I X^J -{1 \over 2} G^{IJ}) \qquad , \qquad 
G_{IJ} = {1 \over 2} \left( \ba{ccc} {1\over S^2} && \\ 
& {1 \over T^2} & \\ && {1 \over U^2} \ea \right)
\ee
\renewcommand{\arraystretch}{1.7}
where the constants $V_I$ parameterize the $U(1)$ subgroup that has
been gauged and $g$ is the corresponding gauge coupling constant, see
\cite{280}. A black hole solution for this Lagrangian has recently be
found \cite{010} and reads
\be420
\ba{l}
ds^2 = - e^{-4U} f dt^2 + e^{2U} \left[ {dr^2 \over f} + r^2 d\Omega_3
\right] \quad, \quad X^I = {e^{2U} \over H_I} \quad , \quad 
A^I = {1 \over H_I} dt \\
e^{6U} = H_1 H_2 H_3  \qquad , \qquad f = 1 + g^2 r^2 e^{6U} \qquad , \qquad
H_I = h_I + {Q_I \over r^2}
\ea
\ee
where $Q_I$ are the different electric charges of the black hole and
the constant parts of the harmonic functions are related to
the vector parameterizing the gauged subgroup: $h_I = 3 V_I$.
If all charges vanish we find the $AdS$ vacuum ($e^{2U} =1$).  The
asymptotic geometry is $R \times S_3$ ($R$ is the time), but it
is straightforward to replace the $S_3$ with a more general manifold
with constant curvature $k$ by\footnote{This generalization
certainly solves the equations of motion, e.g.\ it has been
employed for cosmological solution with general spatial curvature in
\cite{290}. However we did not checked the supersymmetry variations and 
significant modifications
may occur for negative $k$, which however do not change our
subsequent discussion.}
\be430
d\Omega_3 \rightarrow d \Omega_{3,k} = d\chi^2 + \Big({\sin \sqrt k 
\chi \over \sqrt k}\Big)^2 \Big(d \theta^2 + \sin^2\theta d\phi^2
\Big) \quad , \quad f \rightarrow k + g^2 r^2 e^{6U} \ .
\ee
As explained before, the number of non-vanishing charges is related to
the number of unbroken supersymmetries. The single black hole should
be related to $N$=2 super Yang Mills, whereas the double and triple
charged case should describe $N$=1 super Yang Mills. 

In order to keep the expressions simple let us identify all non-vanishing
charges, i.e.\  we write
\be440
e^{6U} = H^n = \Big(1 + {Q \over r^2}\Big)^n
\ee
where $n=1,2,3$ counts the number of equalised charges. Introducing a
new radius by
\be450
\rho^2 = r^2 + Q
\ee
the metric becomes
\be460
ds^2 = e^{2V} dt^2 + e^{-2V} {d\rho^2 \over \Delta} + \rho^2 
\Delta d \Omega_{3,k}  \qquad , \qquad 
e^{2V} = k \Big(1 - { Q\over \rho^2}\Big)^{2n\over 3} + g^2 \rho^2
\Delta 
\ee
with $\Delta = (1-{Q\over \rho^2})^{3-n \over 3}$. As discussed in
\cite{010} the solution for $k=1$, i.e.\ eq.\ (\ref{420}), is
ill-defined.  For the three charges case ($\Delta =1$) it has a naked
singularity and for one or two charges there is a singular
horizon. The best one can get is single charge case where the horizon
is infinitely far away (we keep the asymptotic time, see
introduction). The situation becomes better for $k=-1$, where a zero
of $f$ or $e^{2V}$ indicates a horizon, but in this case the spatial
geometry is given by an hyperboloid.  Furthermore, interesting to note
is the case where the asymptotic space is Minkowskean, i.e.\ $k=0$,
we get back the $AdS$ vacuum for three equal charges (see
(\ref{460})) and only if the charges are not equal one finds
a deformation of the $AdS$ space.

Let us come back to our original motivation, i.e.\ to get a
supergravity picture for the figure \ref{beta}. By comparing our
supergravity solution with the field theory expectations we
immediately run into a contradiction, namely we are interested in a
model that is UV free. The UV region of the field theory translates
into the asymptotic supergravity solution, but all scalar fields are
simply constant there and do not vanish. We may ``cure'' this by
setting some of the constants in the harmonic functions ($h's$) to
zero, but then we are loosing the asymptotic $AdS$ space. As solution
to this problem we will do the same as in the $AdS_3$ case: we look
for the linear dilaton vacuum, which can directly be translated into
field theory.  In the domain wall picture, the linear dilaton vacuum
should describe the second asymptotic region, see figure 2.

Consider the single charge case\footnote{This is not the 
Reissner-Nordstr\"om case as one may expect; Reissner-Nordstr\"om
type solution is obtained by equalising all three charges.} where the
the metric and scalars read
\be470
\ba{l}
ds^2 = - {1 \over H^{2/3}} \, f \, dt^2 + H^{1/3} \Big({dr^2 \over f} + 
	r^2 d\Omega_{3,k} \Big) \quad , \quad H = h + {Q \over r^2} \\
S = H^{-2/3} \quad , \quad T = U = H^{1/3} \quad , \quad
f = (k + g^2 Q + g^2r^2 h) \ .
\ea
\ee
Notice the constraint $STU =1$ eliminates one scalar, so that there is
only one physical scalar either $S$ or $T=U$. Both cases have a
different physical interpretation, in one case the gauge field comes from
an electric 10-d gauge field whereas the other case comes from a
magnetic gauge field (that has been dualized in 5 dimensions).  Lets
start with the magnetic case, so we dualize the electric gauge field
into an antisymmetric tensor and interprete $S$ as our physical scalar
(i.e.\ we have to replace $T$ and $U$ in the action by $T=U=1/\sqrt
S$).  Next, we identify this scalar with the dilaton and
find for the string metric
\be480
ds_{str}^2 = e^{{4 \over 3} \phi} ds^2 =
 -f dt^2 + H \left( {dr^2 \over f} + r^2 d\Omega_{3,k} \right)
\quad , \quad S^{-1} = e^{{4 \over 3} \phi} = 
\Big(h + {Q\over r^2}\Big)^{2/3} \ .
\ee

This is a NS5-brane wrapping the 5-d internal space. In order
to reach the linear dilaton vacuum we have to enter the throat region,
e.g.\ by considering a large charge $Q$ or equivalently
turn off the constant part in the harmonic functions
(i.e.\ $H= { Q \over r^2}$). Defining a new radius we obtain
\be490
ds^2 = - (k + g^2 Q) \, dt^2 +  {Q \over k + g^2 Q} \Big({dr \over r}\Big)^2
 + Q \, d\Omega_{3,k} \qquad , 
\qquad e^{2\phi} =  {Q \over r^2} \\
\ee
and hence for $k=0$ we get the flat space background with a linear
dilaton. How about the dual case, which should be related to a
compactified fundamental string?  In this case we take $T=U$ as
physical scalar and do not dualize the gauge field. As result one
gets in the string frame
\be500
\tilde ds_{str}^2 = e^{{4 \over 3} \phi} ds^2 =
 -{f\over H} dt^2 + \left( {dr^2 \over f} + r^2 d\Omega_{3,k} \right)
\quad , \quad T = e^{-{2 \over 3} \phi} = 
\Big(h + {Q\over r^2}\Big)^{1/3} \ .
\ee
Again, neglecting  the constant part $h$ we find the metric and
dilaton
\be510
\tilde ds^2 = r^2 \left[ - (g^2 + {k \over Q}) \, dt^2 +
d\Omega_{3,k} \right] + {dr^2 \over k + g^2 Q}
\qquad , \qquad e^{2 \phi} = {r^2 \over Q}
\ee
which is again flat space for $k=0$, but the dilaton has been inverted
in comparison to the case before.

These solutions describe the situation as shown in figure
1(a) and 1(b). For the magnetic case as described by (\ref{490}) the
coupling constant vanish asymptotically ($\phi \rightarrow - \infty$
for $r \rightarrow \infty$) and approaching the core of the solution,
i.e.\ moving towards the IR region, we enter the strongly coupled
region ($\phi \rightarrow \infty$ for $r \rightarrow 0$). This is the
situation of figure 1(a). On the other hand the trajectory of figure
1(b) corresponds to the electric case as given in (\ref{510}).  In the
asymptotic vacuum (UV region) we are in strongly coupled region and
approaching the core of the model (IR region) it becomes weakly
coupled ($e^{2\phi} \rightarrow 0$). 

Of course, one would like to have a similar situation as in the
$AdS_3$ case, where we had two dual solutions and could identify them
as the two sides of a domain wall. Here in $AdS_5$, in addition to a
symmetry transformation we had to neglect the constant part in the
harmonic function. It would be very nice, if one could achieve also this 
by a symmetry transformation, perhaps along the line of ungauged case as
discussed in \cite{300}, \cite{310}. 


\section{Conclusions}


Employing the AdS/CFT correspondence we made an attempt to find a
supergravity setup for the RG flows as shown in figure \ref{beta}. In
this picture the asymptotic $AdS$ configuration translates into the UV
region of the worldvolume field theory.  We argued, that a BPS black
hole with a regular horizon translates into the field theory in an
addition IR fixpoint, i.e.\ the near-horizon CFT is dual to the
worldvolume field theory near the the IR fixpoint.  Especially we
discussed the cases of $AdS_3$ and $AdS_5$.

For $AdS_3$ we considered a domain wall solution, that interpolates
between an asymptotic $AdS$ and a linear dilaton vacuum.  On the $AdS$
side it is the BTZ black hole and on the other side it is the $T$-dual
configuration. The CFT corresponding to the asymptotic vacuum differs
from the CFT realized near the horizon. This example should describe
the case shown in figure \ref{beta}(c).

For the $AdS_5$ case we discussed examples that describe the situation
as shown in figure 1(a) and 1(b). The case (a) is described by a single
charged $AdS$ black hole, where the charge comes from a compactified
NS5-brane. The case (b) is the dual electric case, i.e.\ the charge
comes from a fundamental string.  Like for $AdS_3$, for this
interpretation we had in both cases to find a flat space description
(i.e.\ the throat region of the 5-brane). This was done by going into
the string frame and  neglecting the constant part in the harmonic
function.  

The blowing up of the couplings, either in the UV or IR, can be traced
back to the singular horizon of the $AdS$ black hole (\ref{470}) and
indicates that the models are well defined only in certain
regions. One may try to improve the situation by adding more charges,
but the opposite happens. For the single charge case the singular
horizon was infinitely far away, but adding a second charge the
singular horizon is at finite distance. This would translate into the
RG that the coupling would diverge for a finite value of the RG
parameter.  In the case of 3 charges the singular horizon dissappears
but instead we face a naked singularity (also at finite distance). So,
for all these cases the gauge theory should be ill defined and may
wonder about the reason. It is very tempting to speculate that the
singularities for the double and triple charged black hole are related
to anomalies of the gauge theory. Note, in comparison to the $AdS$
vacuum these solutions have only 1/4 of remaining supersymmetries and
hence they correspond to $N$=1 super Yang Mills. In contrast, the
single charged black hole as discussed before translates into $N$=2
super Yang Mills. 

What options do we have to cure this break down?  First, one can add
higher derivative terms, including e.g.\ the Gauss-Bonnet term. E.g.\
in analogy to the $AdS_3$ case one can discuss the pure gravity sector
in terms of a 5-d Chern-Simons theory. This model contains higher
curvature terms and the corresponding black holes are regular, see
\cite{330}. We discussed only a pure bosonic supergravity background,
a further option would be to include also their fermionic super
partners \cite{360}. Since there is no supersymmetry enhancement they
will give non-trivial corrections.  Finally, the inclusion of rotation
may have some attractive features, e.g.\ one can break supersymmetry
already at zero temperature and as claimed in \cite{340}, \cite{350}
$AdS$ gravity allows for rotating BPS black holes with a regular
horizon.


\bigskip \bigskip

{\bf Acknowledgements}

I would like to thank Andreas Karch for extensive discussions
and Arkady Tseytlin for helpful comments.
Research supported by Deutsche  Forschungsgemeinschaft (DFG).

\medskip


\end{document}